\documentclass[11pt,a4paper]{article}
\usepackage{jheppub}
\usepackage{setspace,braket}
\usepackage{amsmath, amssymb, bm}
\usepackage{etoolbox}

    \makeatletter
    \patchcmd{\maketitle}{\@fpheader}{}{}{}
\makeatother
\def\d#1#2{\frac{\displaystyle #1}{\displaystyle #2}}
\def\be{\begin{equation}}
\def\ee{\end{equation}}
\def\bea{\begin{eqnarray}}
\def\eea{\end{eqnarray}}
\def\no{\nonumber}
\def\pt{\partial}

\def\({\left(}
\def\){\right)}
\def\[{\left[}
\def\]{\right]}

\numberwithin{equation}{section}
\renewcommand{\thesection}{\arabic{section}}

\begin{document}
\renewcommand{\thefootnote}{\fnsymbol{footnote}}

\title{Phase transitions of GUP-corrected charged AdS black hole}
\author[a,b]{Meng-Sen Ma, \footnote{The corresponding author, E-mail address: mengsenma@gmail.com}}
\author[b]{Yan-Song Liu,}
\affiliation[a]{Institute of Theoretical Physics, Shanxi Datong
University, Datong 037009, China}
\affiliation[b]{Department of Physics, Shanxi Datong
University, Datong 037009, China}

\abstract{

We study the thermodynamic properties and critical behaviors of the topological charged black hole in AdS space under the consideration of the generalized uncertainty principle (GUP). It is found that
only in the spherical horizon case there are Van der Waals-like first-order phase transitions and reentrant phase transitions. From the equation of state we find that the GUP-corrected black hole can have one, two and three apparent critical points under different conditions. However, it is verified by the Gibbs free energy that in either case there is at most one physical critical point.

}
\maketitle
\onehalfspace

\renewcommand{\thefootnote}{\arabic{footnote}}
\setcounter{footnote}{0}
\section{Introduction}
\label{intro}

Since Hawking-Page phase transition of Schwarzschild-AdS black hole was explored in \cite{Hawking.1983}, phase structures and critical behaviors of various black holes in AdS space have been extensively studied\cite{Chamblin.1999,Chamblin.1999b,Peca.1999,Wu.2000,Myung.2008,Quevedo.2008,Cadoni.2010,Liu.2010,Sahay.2010,Banerjee.2011,Ma.2014,Ma.2014b}. Following \cite{Kastor.2009,Dolan.2011}, the cosmological constant was considered as the thermodynamic pressure and the conjugated quantity was taken as the thermodynamic volume. In this extended phase space, the black hole mass $M$ should be identified with the enthalpy. Although in \cite{Chamblin.1999,Chamblin.1999b}, the Van der Waals(VdW)-like first-order phase transition was first found in the RN-AdS black hole, in the extended phase space it was found the critical behaviors of the RN-AdS black hole have more similarities to that of the VdW liquid/gas system\cite{Kubiznak.2012}. This finding aroused many relevant studies on the critical phenomena of various AdS black holes in the extended phase space\cite{Cai.2013,Chen.2013,Hendi.2013,Altamirano-2014,Mo.2014,Xu.2014,Ma.2015,Ma.2015b,ZhaoHH.2015}. Furthermore, some special critical behaviors such as the reentrant phase transition(RPT), the triple critical point, the isolated critical point and even the ``critical curve" for several black holes have been explored\cite{Altamirano.2013,Wei.2014,Dolan.2014,Frassino.2014,Ming.2017,Sheykhi.2018,Hennigar.2017a,Hennigar.2017b,Dykaar.2017,Ma.2017}.

After considering quantum gravity effects, thermodynamic quantities of black holes may be modified. For example, the generalized uncertainty principle (GUP) will lead to the corrected temperature and entropy\cite{Adler.2001,Medved.2004,ZR.2006,Nouicer.2007,Kim.2008,Majumder.2011,Yang.2016,Vegenas.2017}. Thus, the GUP should also influence the critical behaviors of black holes correspondingly. In \cite{Nouicer.2012}, the author studied the effects of the GUP to all orders in the Planck length on the thermodynamics and the phase transition of the Schwarzschild black hole. In this paper we consider the  usually used more simpler form of the GUP
\be
\Delta x \geq \d{\hbar}{\Delta p}+\d{\alpha^2}{\hbar}\Delta p \geq 2\alpha \sim l_p,
\ee
where $l_p$ is the Planck length, and $\alpha$ is a positive constant with length dimension whose upper limits can be given by the recent discovered gravitational waves\cite{Yang.2017}. On the basis of this relation, the corrected temperature and entropy for some static and stationary black holes were given in \cite{Xiang.2009}. Using these corrected thermodynamic quantities, we have studied the critical behaviors of the Schwarzschild-AdS black hole and the RN-AdS black hole in \cite{EPL.2018}. With the GUP corrections, we find that the Hawking-Page phase transition for the AdS black holes no longer always occurs. In this paper, we will further study the critical behaviors and phase transitions of the corrected charged topological AdS black hole in the extended phase space. We find that a combination  of $\alpha$ and the electric charge $Q$ can be used to classify the various kinds of critical behaviors.

The plan of this paper is as follows:
In Sec.2 we introduce the corrected thermodynamic quantities of the charged AdS black hole and simply discuss their properties.
In Sec.3 we find the critical points and analyze the numbers of the critical points.
In Sec.4 we study the critical behaviors of the black hole according to the Gibbs free energy.
In Sec.5 we summarize our results and discuss the possible future directions.

\section{Thermodynamics of the charged topological AdS black hole with GUP correction}

In Einstein gravity in four dimensional spacetime, we have the charged topological AdS black hole solution,
\be
ds^{2}=-f(r)dt^{2}+f(r)^{-1}dr^{2}+r^{2}d\Omega_k^2,
\ee
with the metric function\cite{cai.1999}
\be\label{metric}
f(r)=k-\d{8\pi G M}{\Sigma_k r}+\d{16\pi^2G^2Q^2}{\Sigma_k^2 r^2}+\d{r^2}{l^2},
\ee
where the parameters $M,~Q$ are the ADM mass and electric charge of the black hole and $l$ represents the cosmological radius. $d\Omega_k^2$ denotes the line element of a two-dimensional Einstein space with constant scalar curvature $2k$ and volume $\Sigma_k$. Without loss of generality, one can take $k=1$ (spherical horizon), $k=0$ (planar/toroidal horizon), and $k=-1$ (hyperbolic horizon). Besides, we set $4\pi G/\Sigma_k=1$ for simplicity. Although $\Sigma_k$ has different values for different $k$, this simplification will not affect our physical results.

According to the metric function in Eq.(\ref{metric}), the black hole mass is
\be
M=\frac{3 k r_h^2+8 \pi  P r_h^4+3 Q^2}{6 r_h},
\ee
where $r_h$ denotes the position of the event horizon of the black hole. Here $P$ is the thermodynamic pressure and is taken to be $P=-\d{\Lambda}{8\pi}=\d{3}{8\pi l^2}>0$.

The surface gravity of the black hole is
\be\label{kappa}
\kappa=\d{f'(r_h)}{2}=\frac{k r_h^2+8 \pi  P r_h^4-Q^2}{2 r_h^3}.
\ee
In the semiclassical case, the temperature and entropy for the black hole are
\be\label{TS}
T=\d{\hbar\kappa}{2\pi}, \quad  S=\d{A}{4\hbar}.
\ee

As a thermodynamic system, the thermodynamic quantities of the black hole should satisfy the thermodynamic identity:
\be\label{1stlaw}
d M=T d S+\Phi d Q+VdP,
\ee
where the electric potential measured at infinity with reference to the horizon is $\Phi=Q/r_{h}$ and the thermodynamic volume is $V=4\pi r_{h}^3/3$.

Generally, black hole entropy should be a function of the horizon area, namely $S=S(A)$\cite{Bekenstein}. Therefore, the temperature of a black hole can be generally
expressed as\cite{Xiang.2009}
\be
T=\left.\d{\pt M}{\pt S}\right|_Q=\d{d A}{d S}\times \left.\d{\pt M}{\pt A}\right|_Q=\d{d A}{d S}\times \d{\kappa}{8\pi}.
\ee
According to Heisenberg uncertainty principle, one can derive $\d{d A}{d S}\simeq \d{\Delta A}{\Delta S}=const$. This is just the work of Bekenstein and Hawking, which give the results in Eq.~(\ref{TS}).

\begin{figure}
\center{
\includegraphics[width=6cm]{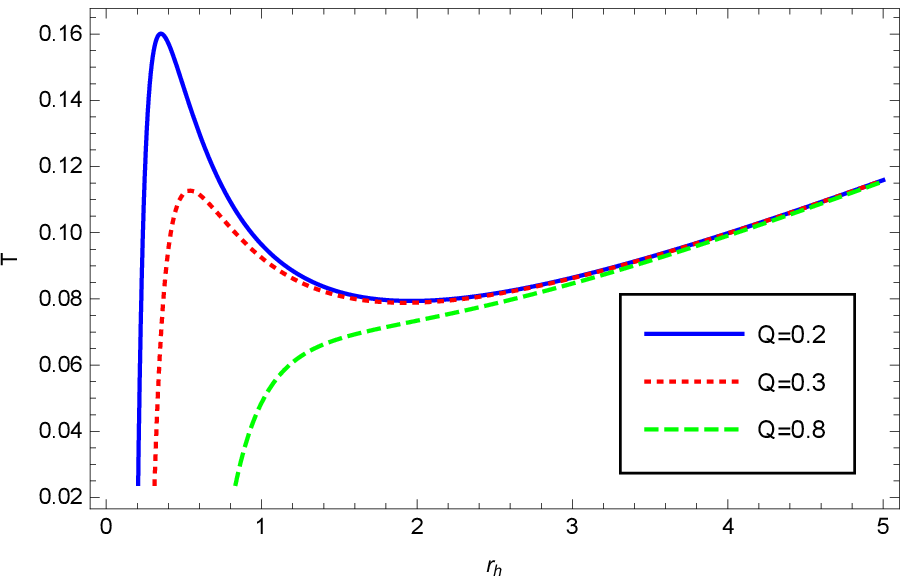}\hspace{0.5cm}
\includegraphics[width=6cm]{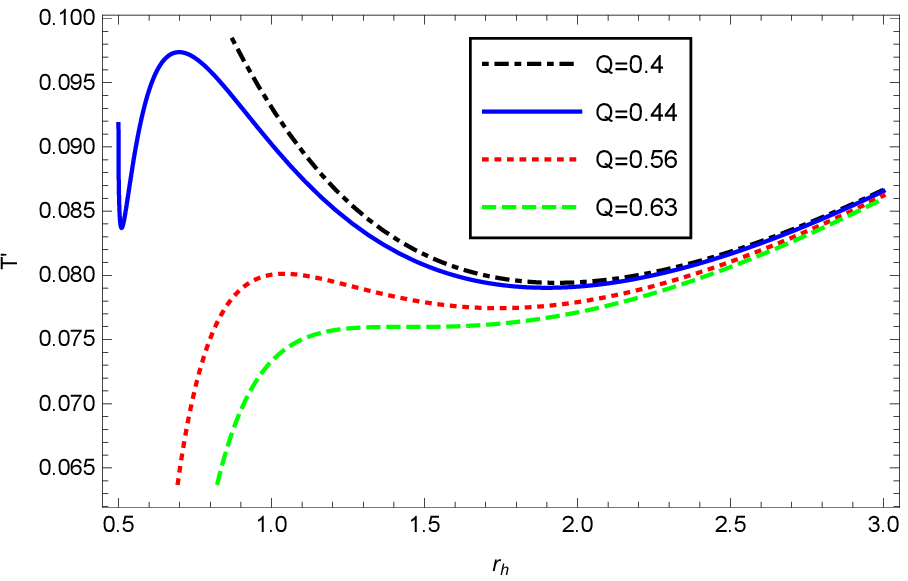}
\caption{The left panel: the standard temperature of the RN-AdS black hole. The right panel: the GUP-corrected temperature with $\alpha=0.5$. In both cases we take $k=1$ and $P=0.01$.}\label{figT}}
\end{figure}

Considering the effect of GUP, it is shown that\cite{Xiang.2009}
\be
\d{d A}{d S}\simeq \d{(\Delta A)_{min}}{(\Delta S)_{min}}=4\hbar',
\ee
where $\hbar'$ is the effective Planck ``constant" and is defined as
\be
\hbar'=\frac{2\hbar}{\alpha^2} \left(r_h^2-r_h \sqrt{r_h^2-\alpha^2}\right).
\ee
Thus, the GUP-corrected black hole temperature becomes
\be\label{CT}
T'=\d{\hbar'\kappa}{2\pi}=\frac{\hbar \left(r_h-\sqrt{r_h^2-\alpha ^2}\right) \left(k r_h^2+8 \pi  P r_h^4-Q^2\right)}{2 \pi  \alpha ^2 r_h^2}.
\ee
From Eq.(\ref{kappa}), one can see that the usual temperature of the charged AdS black hole will become negative for very small $r_h$. While $T'$ give a mandatory requirement $r_h\geq \alpha$, from which we find that the temperature $T'$ can be always positive when the condition $Q^2<\alpha^2(k+8\pi P \alpha^2)$ is satisfied. In Fig.\ref{figT}, we compare the behaviors of the usual temperature $T$ and the corrected temperature $T'$ for the charged AdS black hole. For smaller $Q$, $T'$ is indeed always positive. Besides, with the GUP corrections the $T'-r$ curve exhibits more fruitful structures.

Because GUP only constrains the minimal length, thus only influences the temperature and the entropy. The electric charge and the
electric potential will remain unchanged. The first law of black hole thermodynamics $dM=T'dS'+\Phi dQ+VdP$ should still be established in this case.
Therefore, the GUP-corrected entropy of the black hole can be derived
\bea\label{CS}
S'&=&\left.\int\frac{dM}{T'}\right|_{Q,P}=\left.\int\frac{1}{T'}\frac{\pt M}{\pt r}\right|_{Q,P}dr+S_0 \no \\
  &=&\frac{\pi}{2\hbar}  \left[r_h^2+ r_h \sqrt{r_h^2-\alpha^2}-\alpha^2 \ln \left(\d{\sqrt{r_h^2-\alpha^2}+r_h}{\alpha}\right)\right], \\
  &=&\frac{A}{4\hbar}-\frac{\pi \alpha^2}{4\hbar}\ln\frac{A}{\pi\alpha^2}+\cdots
\eea
Here the effect of GUP leads to a subleading logarithmic term, which also exists in many other quantum corrected entropy. Our entropy is a little different from that in \cite{Xiang.2009}, where the authors take an indefinite integral and treat the integral constant as zero. We take the integration constant $S_0=\alpha^2\ln\alpha$ to obtain a dimensionless logarithmic term.  $S_0$ cannot be fixed by some physical consideration. To determine $S_0$ completely, one has to invoke the quantum theory of gravity. It should be noted that the corrected entropy is independent of the parameter $k$ and it is always positive.
Moreover, due to the existence of the logarithmic term in the corrected entropy, the Smarr formula no more exists.

\section{Multiple critical points}

In this section, we try to ascertain the number of the critical points. Below we always set $\hbar=1$ for simplicity. From Eq.(\ref{CT}), we can derive the equation of state
\be\label{PV}
P=\frac{r_h^2 \left[2 \pi  T' \left(\sqrt{r_h^2-\alpha ^2}+r_h\right)-k\right]+Q^2}{8 \pi  r_h^4}.
\ee
To derive the critical points, one should solve the following two equations
\be
\d{\partial P}{\partial r_{h}}=\d{\partial^2 P}{\partial r_{h}^2}=0.
\ee
One can also use another equivalent pair of equations: $\d{\partial T'}{\partial r_h}=\d{\partial^2 T'}{\partial r_h^2}=0$ to determine the critical points of the system. In either case, the results are the same.

The two expressions are lengthy, we will not list them here. Combining them, we obtain an equation
\bea\label{constraint}
&&r_h^4 \left(12 Q^2-\alpha ^2 k\right)+2 r_h^3 \sqrt{r_h^2-\alpha ^2} \left(\alpha ^2 k+3 Q^2\right)-3 k r_h^5 \sqrt{r_h^2-\alpha ^2}-26 \alpha ^2 Q^2 r_h^2 \\ \nonumber
&&-4 \alpha ^2 Q^2 r_h \sqrt{r_h^2-\alpha ^2}+16 \alpha ^4 Q^2=0.
\eea
We set $\beta=\sqrt{1-\alpha^2/r_{h}^2}$, thus $0\leq\beta\leq1$. Utilizing $\beta$, Eq.(\ref{constraint}) can be simplified to
\be\label{constraint1}
\alpha ^4 (2 \beta +1) \left[\alpha ^2 \left(\beta ^2-\beta +1\right) k+2 \left(4 \beta ^5-\beta ^4-5 \beta ^3+2 \beta ^2+\beta -1\right) Q^2\right]=0.
\ee
In the case of $\alpha \neq 0$, we obtain a constraint equation
\be\label{constraint2}
\frac{k\alpha^2}{Q^2}=\frac{2 \left(4 \beta ^5-\beta ^4-5 \beta ^3+2 \beta ^2+\beta -1\right)}{-\beta ^2+\beta -1}.
\ee

\begin{figure}
\center{
\includegraphics[width=6cm]{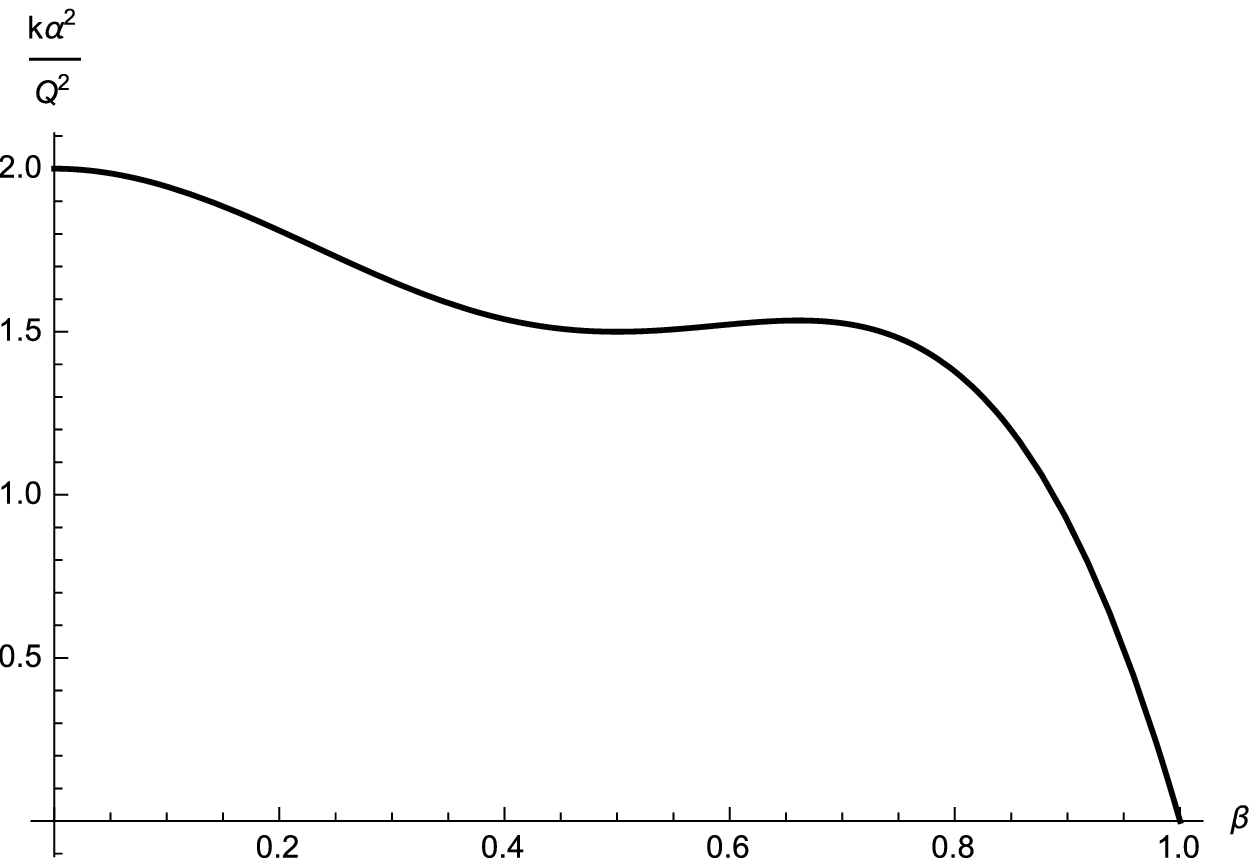} \hspace{0.5cm}
\includegraphics[width=6cm]{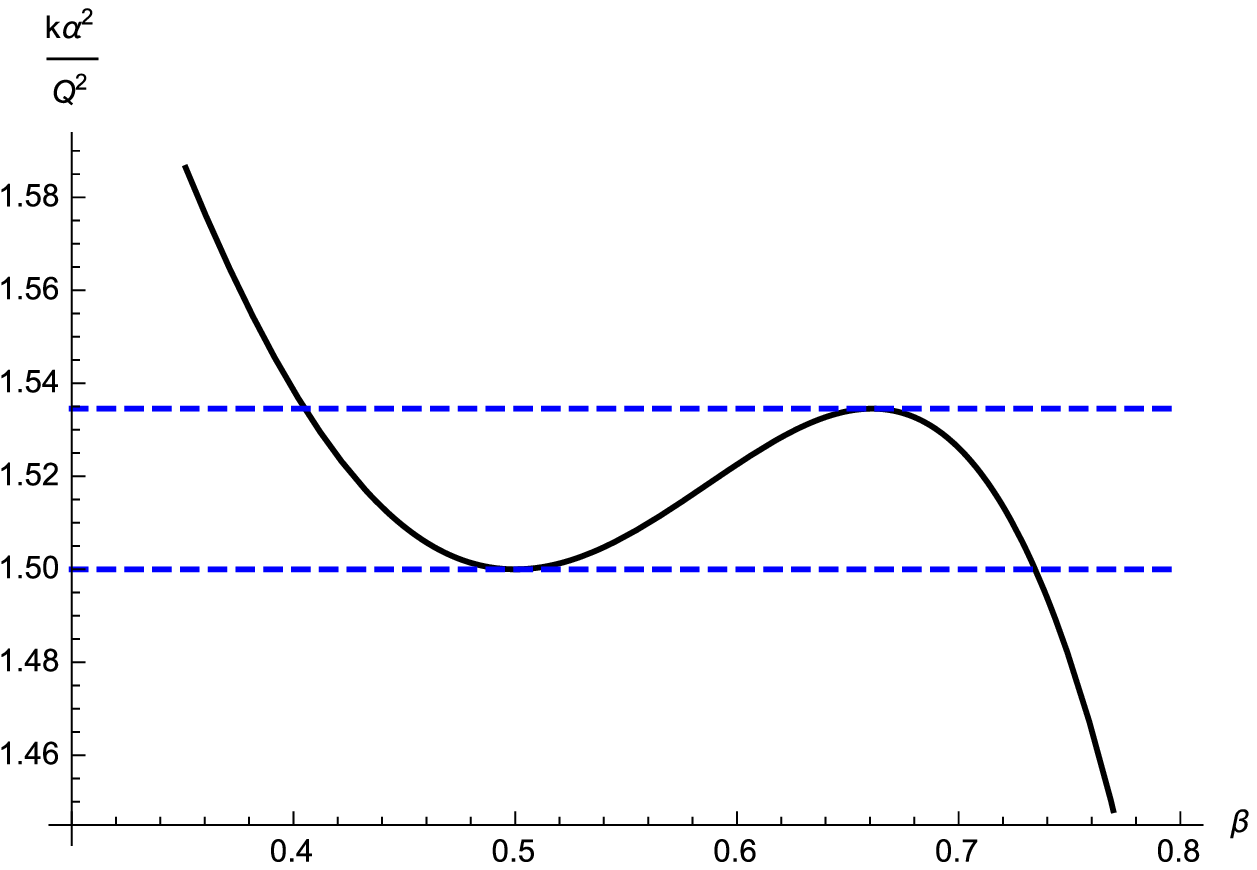}
\caption{Number of the apparent critical points. There are at most three critical points for $1.5<\alpha^2/Q^2<1.535$.}\label{figbeta}}
\end{figure}

As is depicted in Fig.\ref{figbeta}, the right-hand side of Eq.(\ref{constraint2}) is nonnegative\footnote{In fact, when $\beta=1$, the RHS of Eq.(\ref{constraint2}) is zero. However, because we are only interested in the nontrivial case with $\alpha \neq 0$,  this excludes the possibility of $\beta = 1$.}. This means that there is no $P-V$ criticality in the cases with $k=0$ and $k=-1$. Besides, when the electric charge $Q=0$, Eq.(\ref{constraint1}) becomes
\be
k\alpha ^6 (2 \beta +1) \left(\beta ^2-\beta +1\right) =0,
\ee
which also has no real solutions for $\beta$ in the range $0\leq\beta\leq1$. If $k=0$, this equation is well satisfied, however one can easily check that there is still no critical point in the $(k=0,~Q=0)$ case.

Thus, below we are only concerned with the $k=1$ case with nonzero electric charge $Q$.
We find that the number of critical points depends on the value of $\alpha^2/Q^2$. When $\alpha^2/Q^2>1.535$ or $\alpha^2/Q^2<1.5$, there is only one critical point.
When $\alpha^2/Q^2=1.535$ or $\alpha^2/Q^2=1.5$, there are two critical points. And three critical points occur when $1.5<\alpha^2/Q^2<1.535$.

On the basis of these critical points, we can further discuss the heat capacity at constant pressure,
\be
C_p=\left.T'\frac{\partial S'}{\partial T'}\right|_{P,Q}=\left.T'\frac{\partial S'/\pt r_h}{\partial T'/\pt r_h}\right|_{P,Q},
\ee
which can reflect the local thermodynamic stability of the black hole.
The divergence points of the heat capacity occur at zeros of $\partial T'/\pt r_h$, which are the extremal points in the $T'-r_h$ curve. The sign of $C_p$ is also completely determined by $\partial T'/\pt r_h$. Because $\partial S'/\pt r_h=\pi  \left(\sqrt{r_h^2-\alpha ^2}+r_h\right)>0$ and the corrected temperature is greater than zero if $Q^2<\alpha^2(1+8\pi P)$.

\section{The critical behaviors and Gibbs free energy}

Below we discuss the critical behaviors of the RN-AdS black hole according to the numbers of the apparent critical points.

\subsection{One critical point}
\begin{figure}
\center{
\includegraphics[width=6cm]{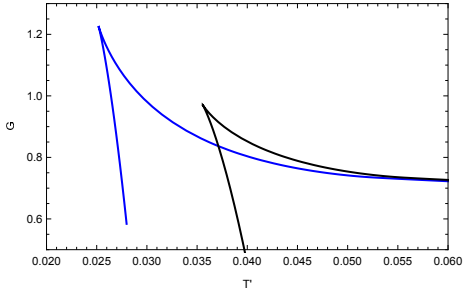}
\caption{The case with $\alpha^2/Q^2=1.6$. The blue curve curve and the black curve correspond to $P=0.001$ and $P=0.002$, respectively.}\label{fig11}}
\end{figure}

In this case, we take $\alpha^2/Q^2=1.4$ and $\alpha^2/Q^2=1.6$, respectively. First, for $\alpha^2/Q^2=1.6$ one can easily find that the critical value of the pressure is negative, which means that no second-order phase transition occurs. In fact, there is also no VdW-like first-order phase transition. As is shown in Fig.\ref{fig11}, the Gibbs free energy exhibits a cusp for any positive given pressure.

\begin{figure}
\center{
\includegraphics[width=6cm]{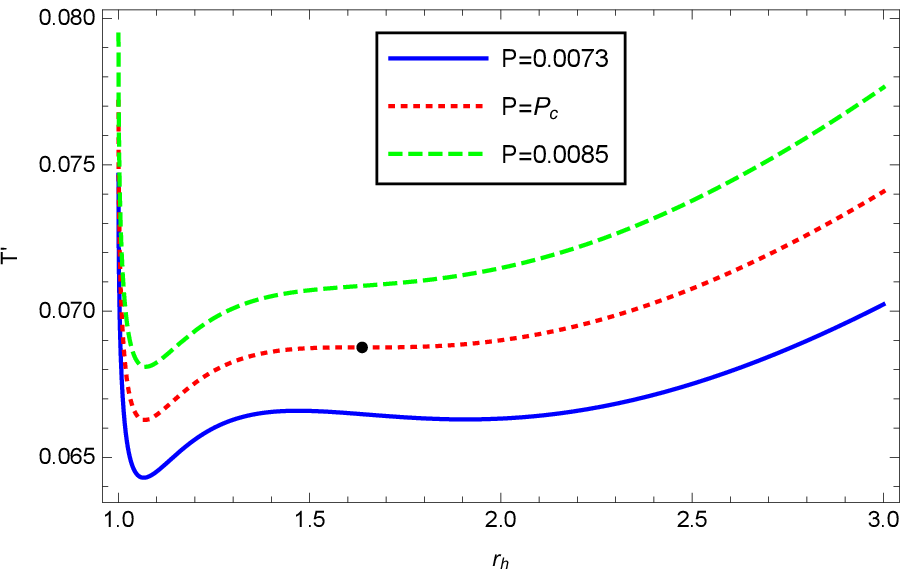}\hspace{0.5cm}
\includegraphics[width=6cm]{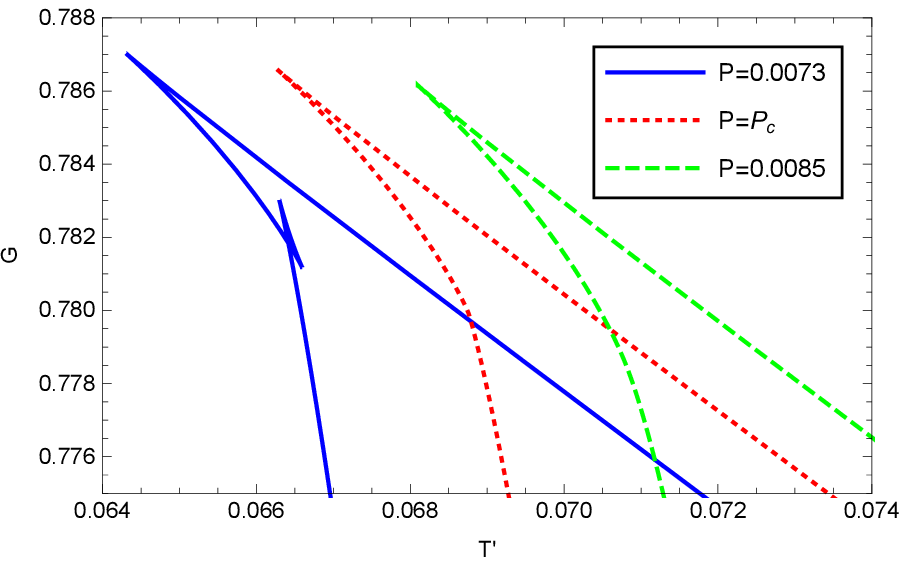}
\caption{The $\alpha^2/Q^2=1.4$ case with $\alpha=1$ and $Q=0.845$. The critical pressure is $P_c=0.0079$.}\label{fig12}}
\end{figure}

\begin{figure}
\center{
\includegraphics[width=5cm]{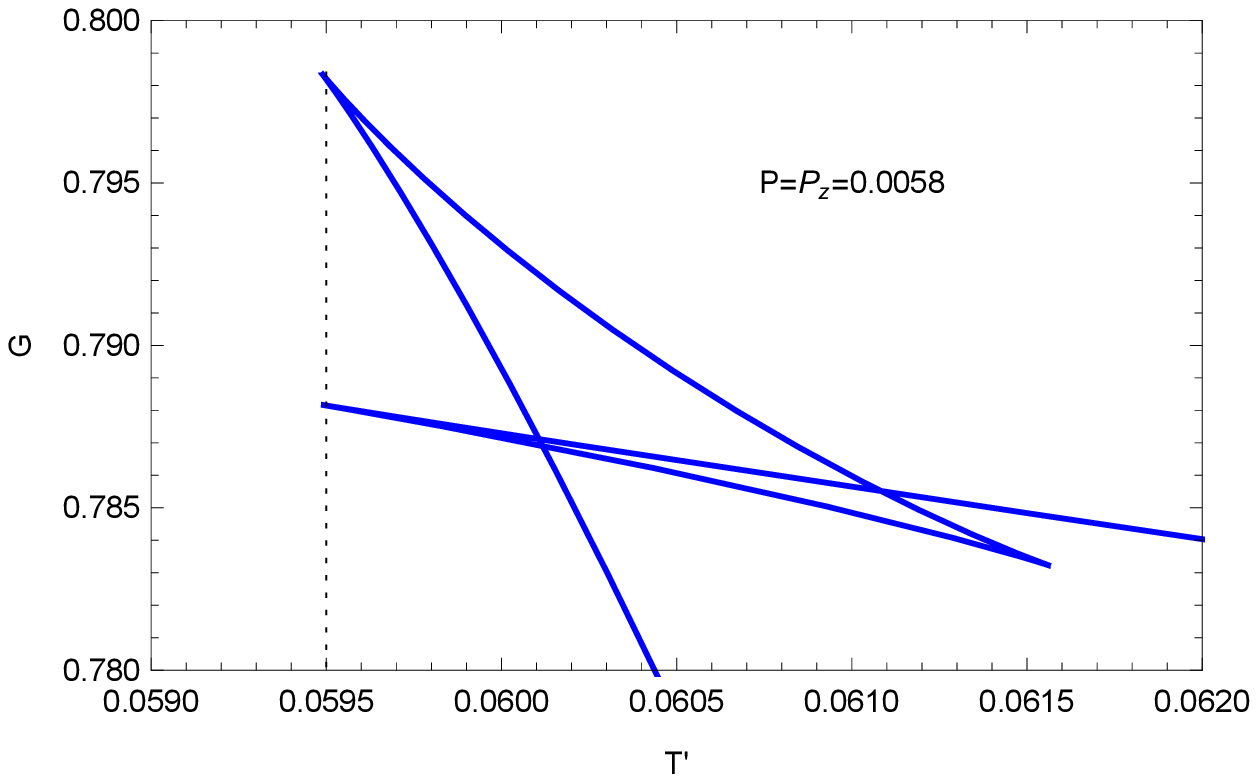}\hspace{0.5cm}
\includegraphics[width=5cm]{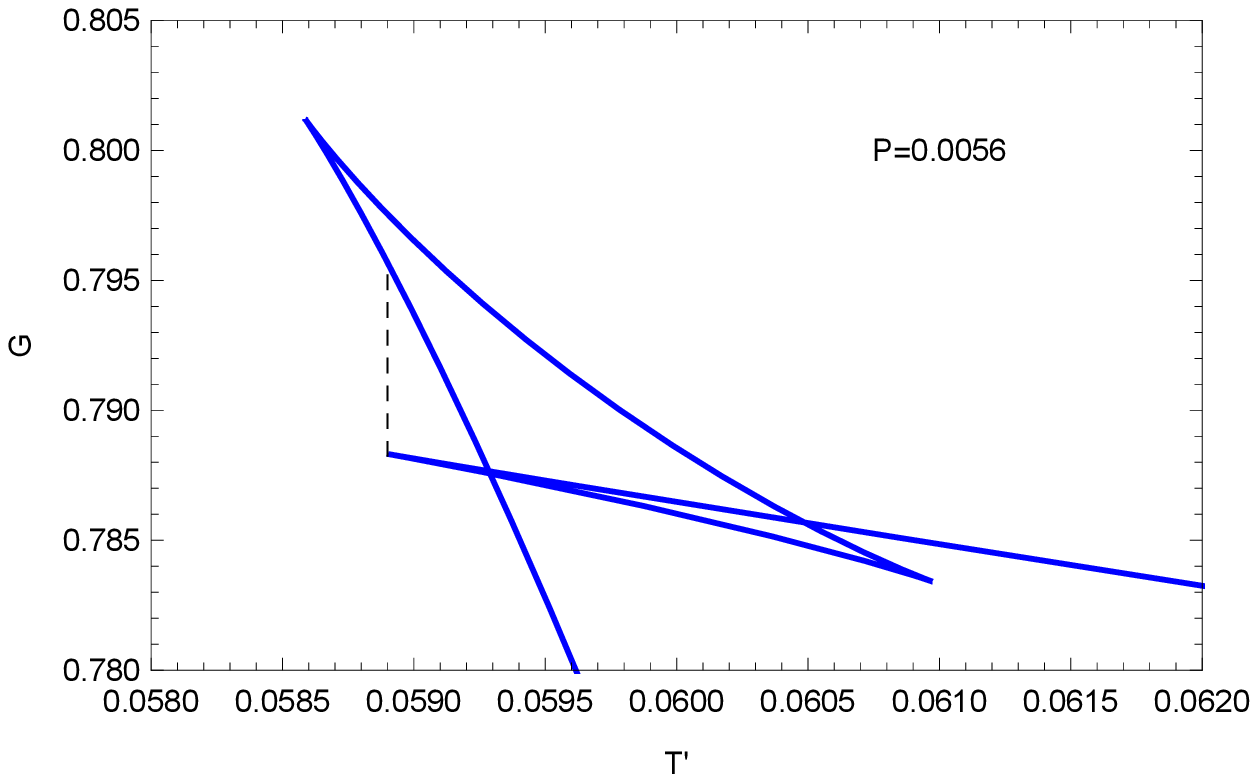}\hspace{0.5cm}
\includegraphics[width=5cm]{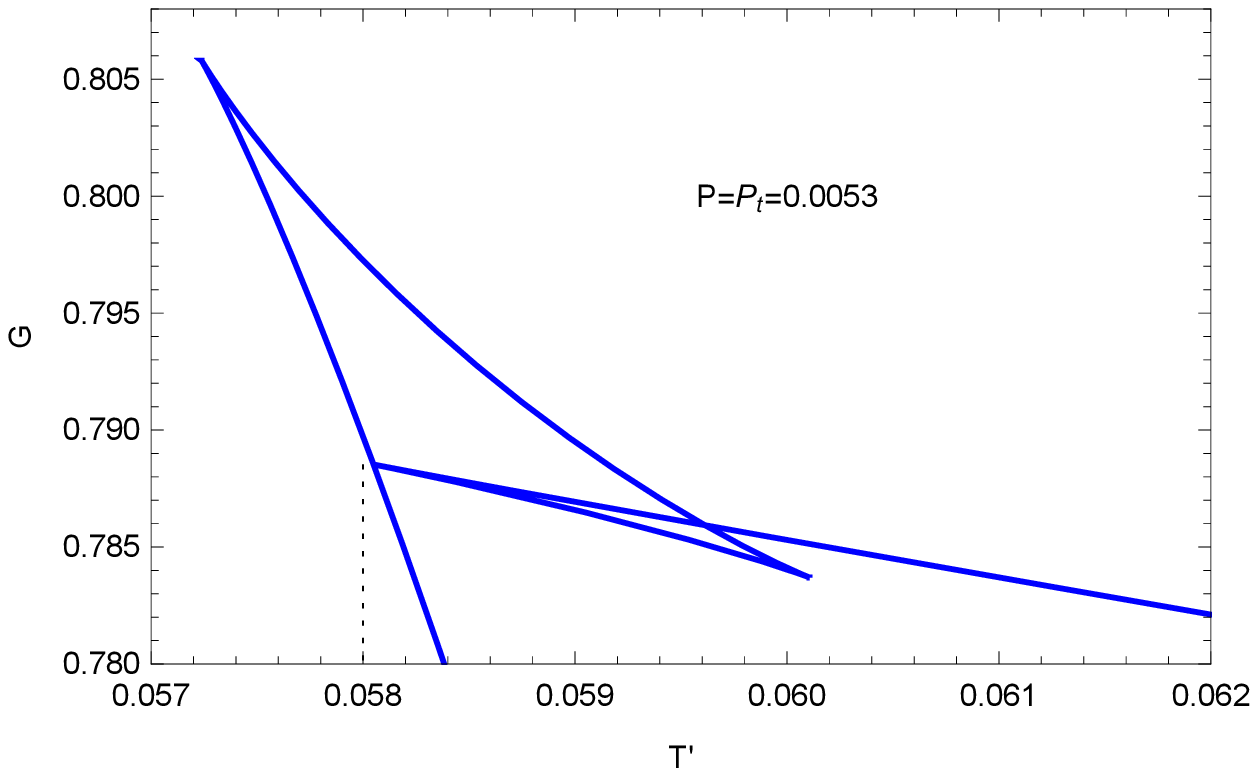}
\caption{Characteristic behavior of Gibbs free energy for the reentrant phase transition. Here we also take $\alpha=1$ and $Q=0.845$.}\label{fig13}}
\end{figure}

The critical behaviors in the case with $\alpha^2/Q^2=1.4$ have been analyzed in \cite{EPL.2018}. When the pressure is greater than the critical pressure $P_c$, there is two branches in the $T'-r_h$ curve, which corresponds to a cusp in $G$. When the pressure is lower than $P_c$, the $T'-r_h$ curve exhibits four branches. From left to right, we call them the small black hole, the left-intermediate black hole, the right-intermediate black hole and the large black hole. According to the slope of the $T'-r_h$ curve one can figure out that $C_p$ is negative in the small black hole branch and the right-intermediate black hole branch and it is positive in the left-intermediate black hole branch and the large black hole branch. From the $G-T'$ figure, one can see a standard VdW-like first-order phase transition. We illustrate this in Fig.\ref{fig12}. For $P\in (P_t,~P_z)$, reentrant phase transition takes place. In this case, if starting off from the largest $G$, the black hole will first evolve along the branch of the large black hole. Then at some point the black hole will undergo a zero-order phase transition and jump to the left-intermediate branch. Finally, undergoing a first-order phase transition, the black hole returns back to the original large black hole. This process has been illustrated in Fig.\ref{fig13}. When $P<P_t$, the large black hole is globally thermodynamic stable, thus no phase transition occurs.

\subsection{Two critical points}

\begin{figure}
\center{
\includegraphics[width=6cm]{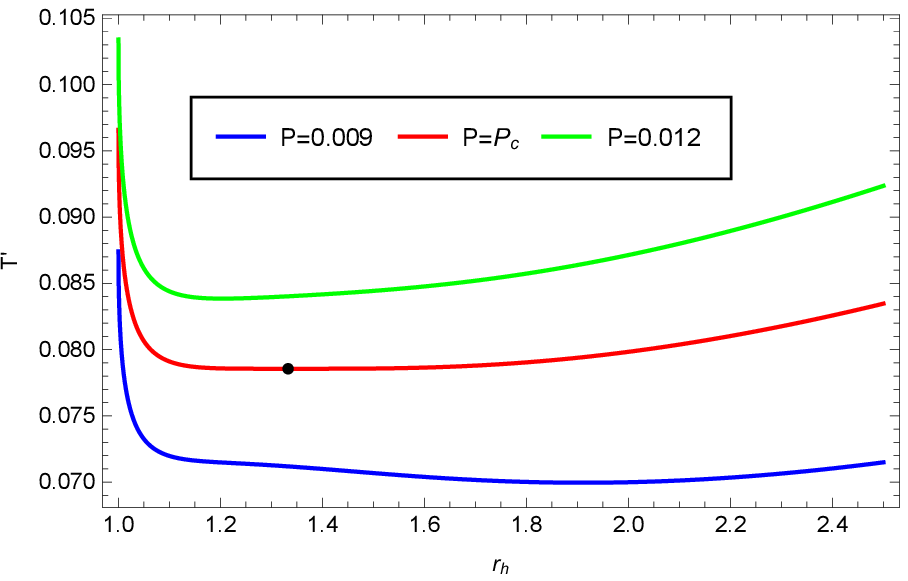}\hspace{0.5cm}
\includegraphics[width=6cm]{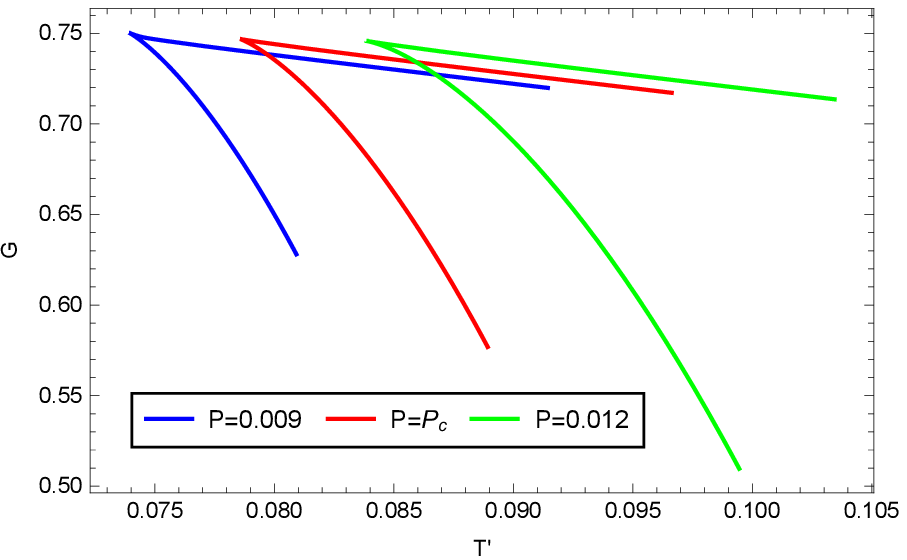}
\caption{The $\alpha^2/Q^2=1.535$ case with $\alpha=1$ and $Q=0.807$.}\label{fig21}}
\end{figure}

When $\alpha^2/Q^2=1.535$, one can easily check that the smaller critical point is false because the critical pressure $P_c$ is negative. The larger critical point is $(P_c=0.0103,~T_c=0.0786,~r_{hc}=1.333)$. However, as is shown in Fig.\ref{fig21}, the black hole only has two branches and exhibits a cusp in $G$. The lower branch has the positive heat capacity and smaller Gibbs free energy. Thus it is more stable for the large black hole and no phase transition occurs in this case.

\begin{figure}
\center{
\includegraphics[width=6cm]{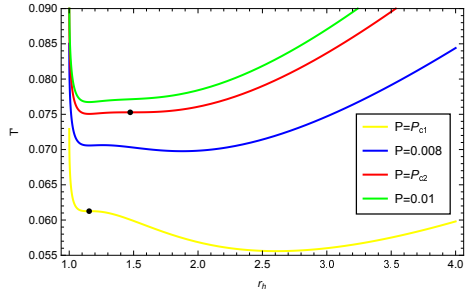}
\caption{The $\alpha^2/Q^2=1.5$ case with $\alpha=1$ and $Q=0.816$. The reentrant phase transition takes place for $P \in (P_{t},~P_{z})$ with $P_{t}=0.00887,~P_{z}=0.009$.}\label{fig22}}
\end{figure}

For $\alpha^2/Q^2=1.5$, the smaller critical point is $(T_{c1}=0.0613,~P_{c1}=0.00497,~r_{c1}=1.155)$, and the larger critical point is $(T_{c2}=0.0753,~P_{c2}=0.00945,~r_{c2}=1.474)$. As is illustrated in Fig.\ref{fig22}, for $P \in (P_{c1},~P_{c2})$ the black hole has four branches. Similar to the one critical point case ($\alpha^2/Q^2=1.4$), for $P \in (P_{t},~P_{c2})$ the black hole always has a VdW-like first-order phase transition and for $P \in (P_{t},~P_{z})$ there exists the reentrant phase transition. Below a physical critical point there should be a VdW-like first-order phase transition. Because $P_{c1}<P_t$, for $P\in (P_{c1},~P_t)$ or $P<P_{c1}$ the large black hole is always globally thermodynamically stable. Therefore, the smaller critical point is indeed an apparent one, which does not correspond to any second-order phase transition. In this case, the Gibbs free energy has the similar behaviors to that in Fig.\ref{fig12} and Fig.\ref{fig13}.

\subsection{Three critical points}

In this case, we select $\alpha^2/Q^2=1.51$. Three apparent critical points are $(T_{c1}=0.0497,~P_{c1}=0.00111,~r_{c1}=1.118)$, $(T_{c2}=0.0717,~P_{c2}=0.00826,~r_{c2}=1.207)$ and $(T_{c3}=0.0761,~P_{c3}=0.00966,~r_{c3}=1.45)$, respectively.

\begin{figure}
\center{
\includegraphics[width=5cm]{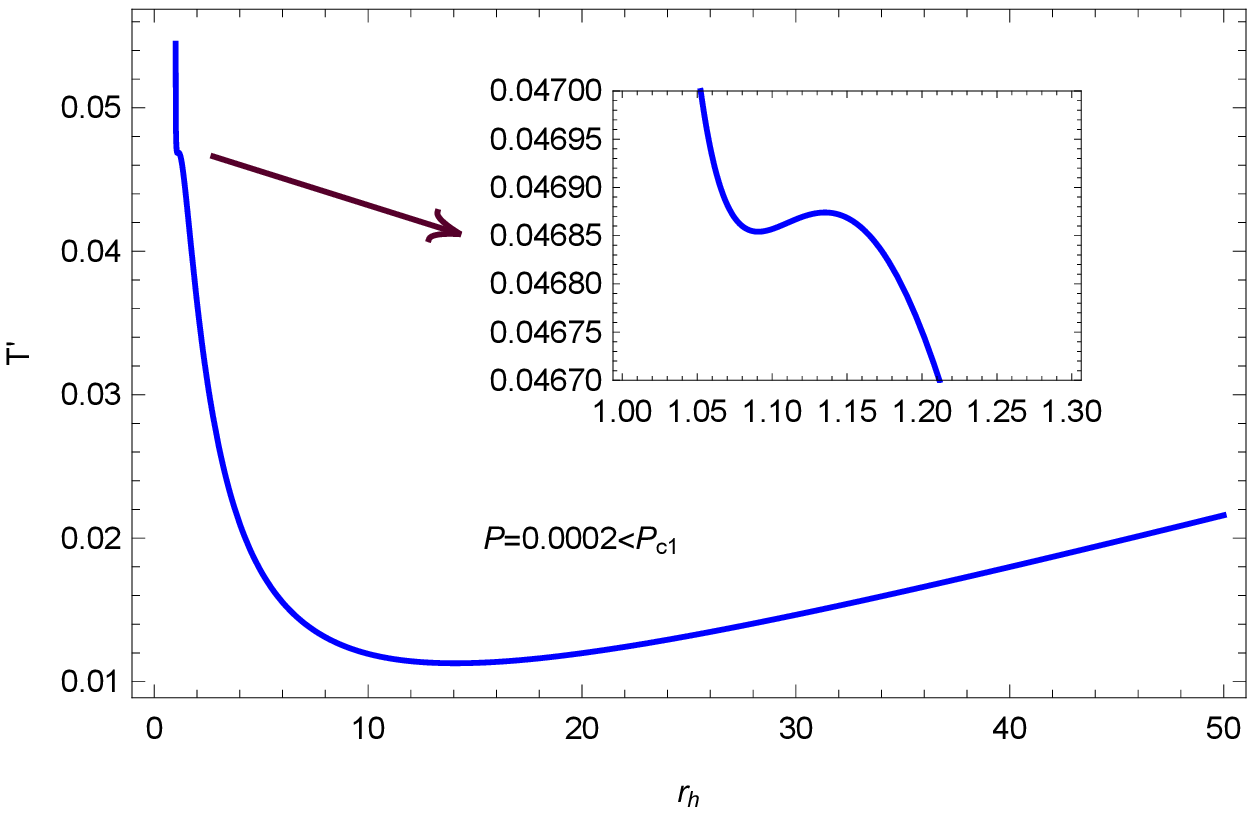}\hspace{0.5cm}
\includegraphics[width=5cm]{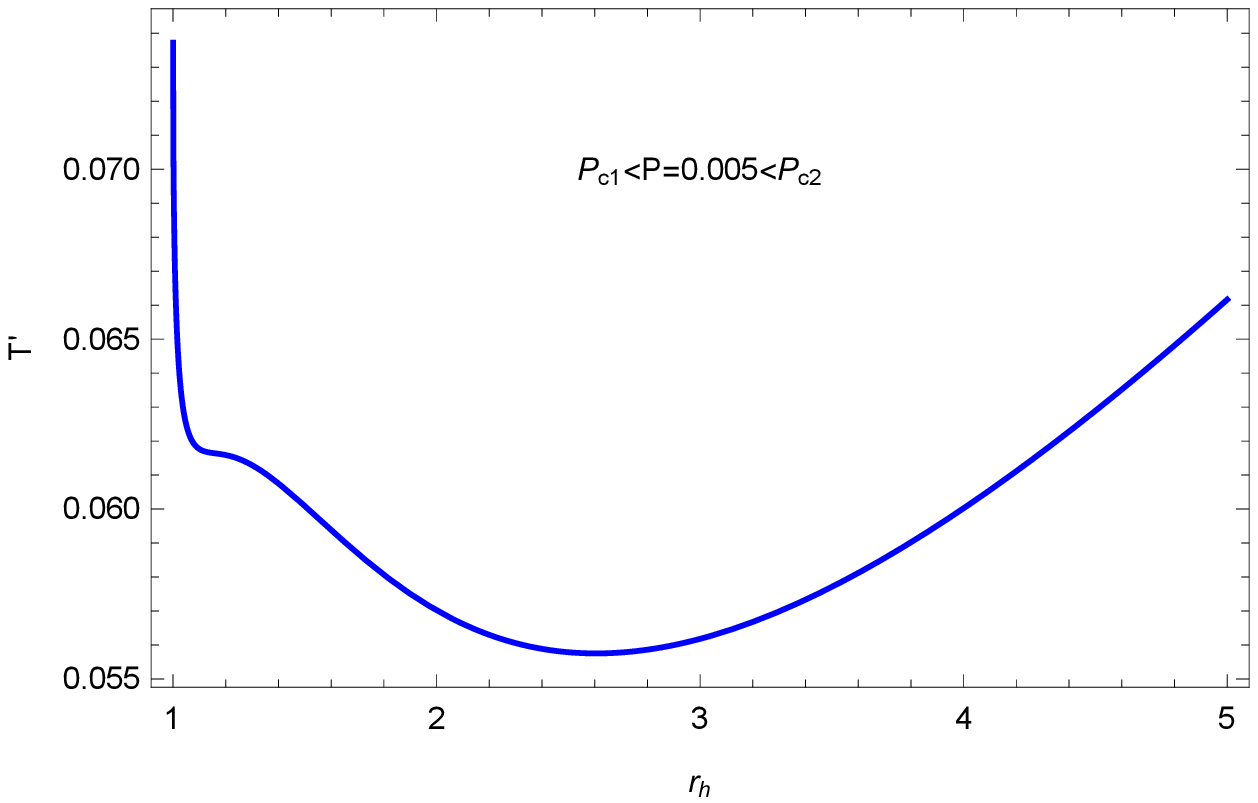}\hspace{0.5cm}
\includegraphics[width=5cm]{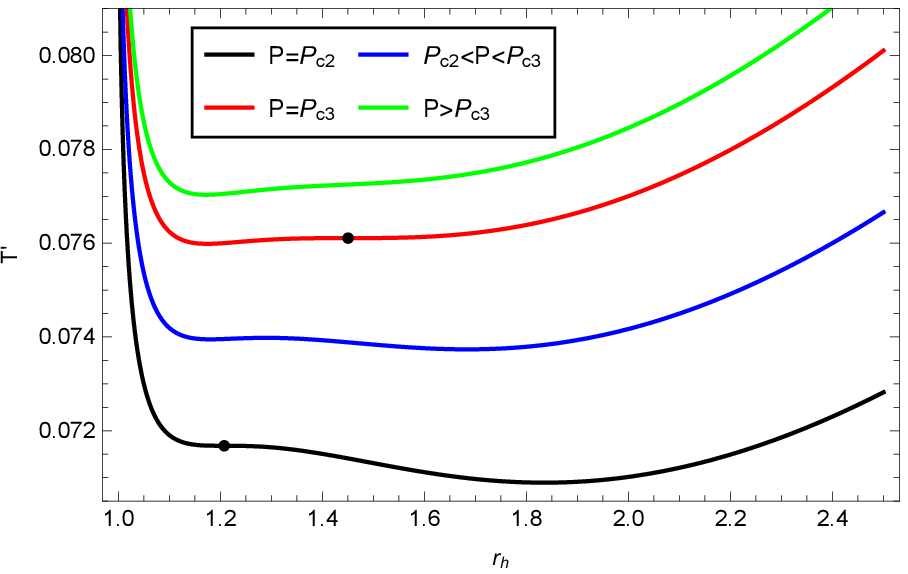}
\caption{$T'-r_{h}$ curves for $\alpha^2/Q^2=1.51$ with $\alpha=1$ and $Q=0.814$.}\label{fig31}}
\end{figure}

\begin{figure}
\center{
\includegraphics[width=5cm]{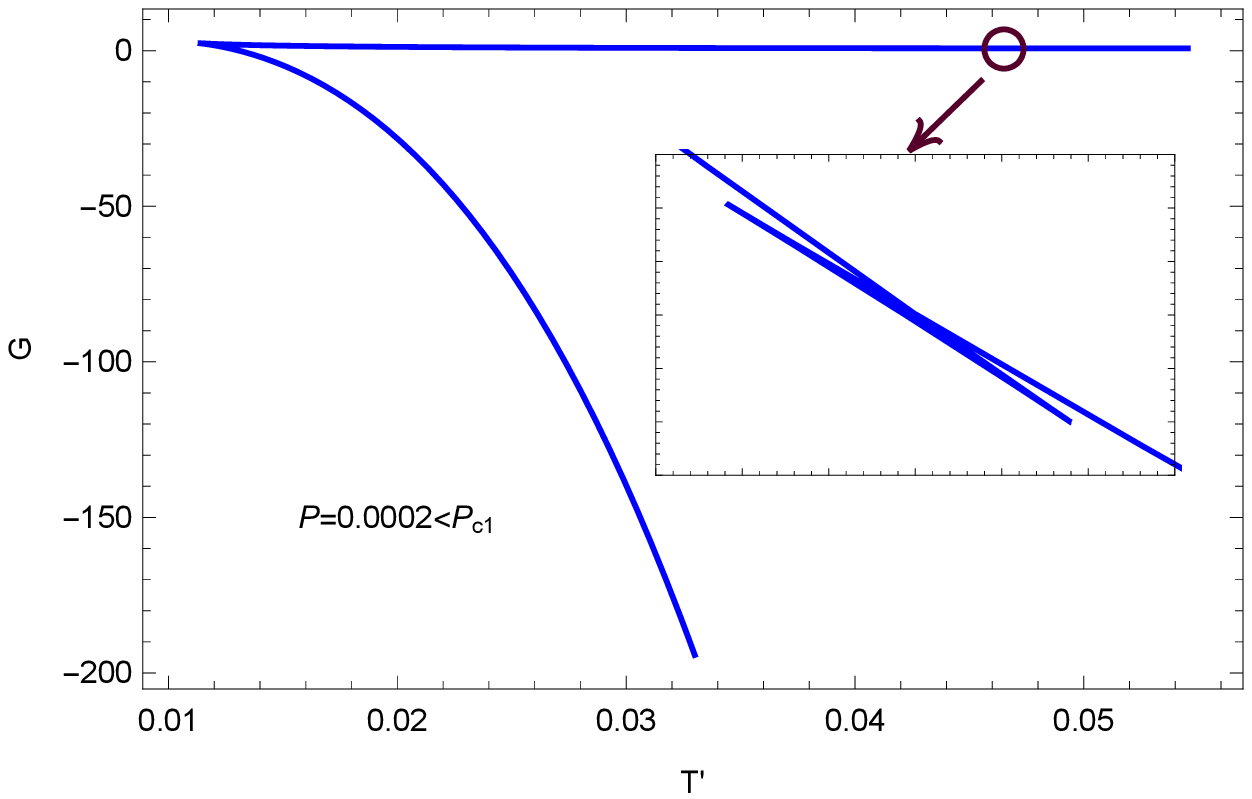}\hspace{0.5cm}
\includegraphics[width=5cm]{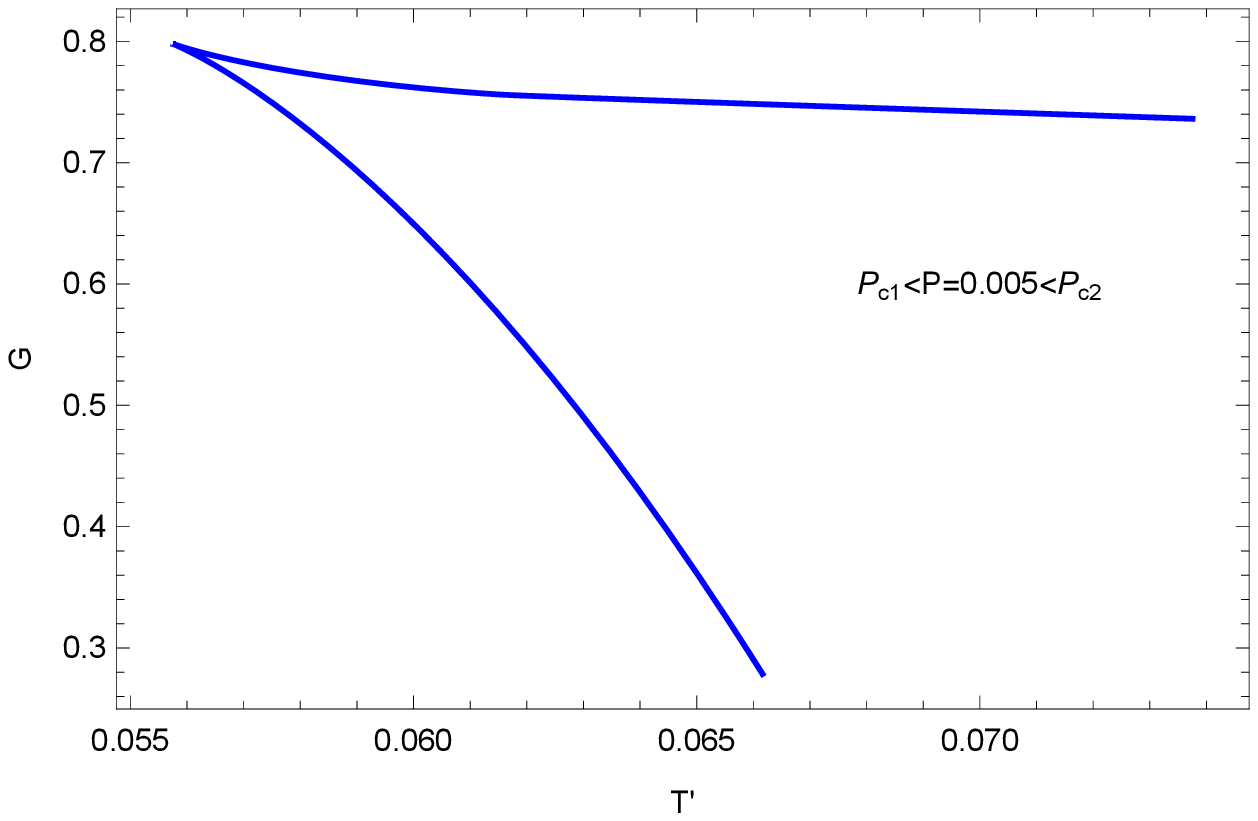}\hspace{0.5cm}
\includegraphics[width=5cm]{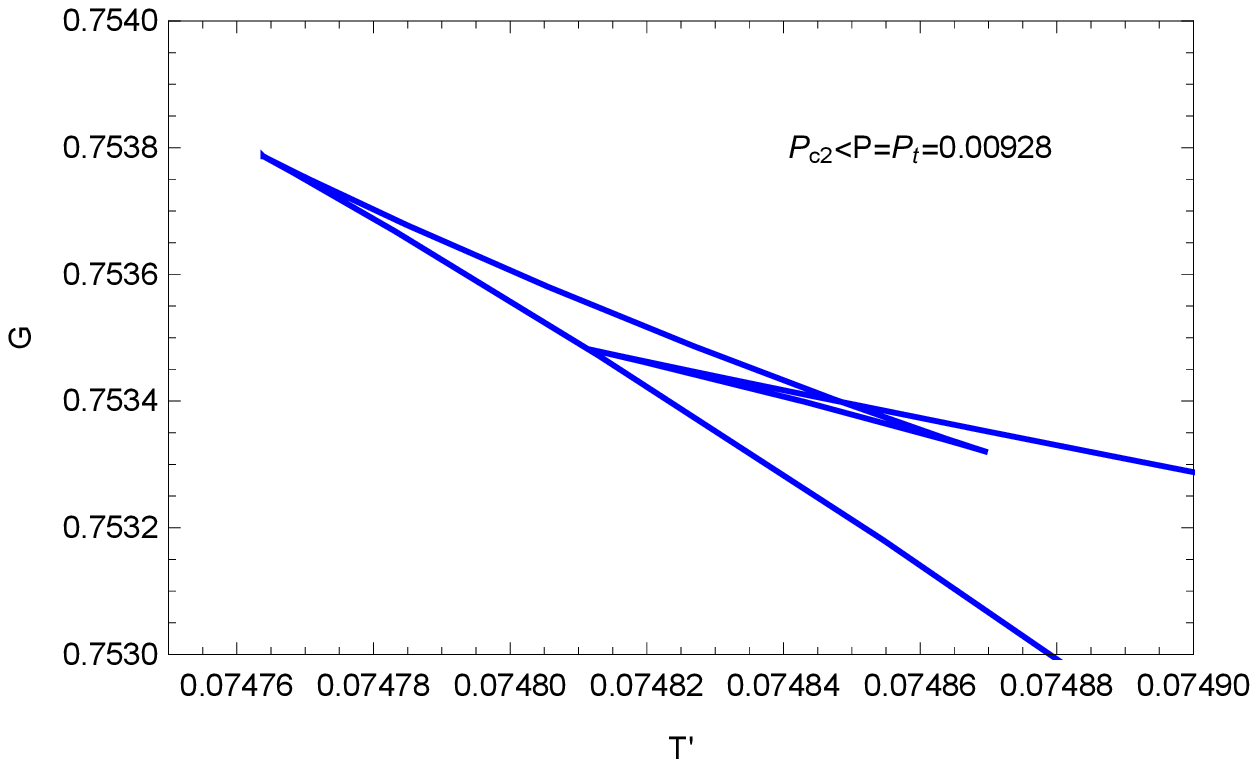} \\
\includegraphics[width=5cm]{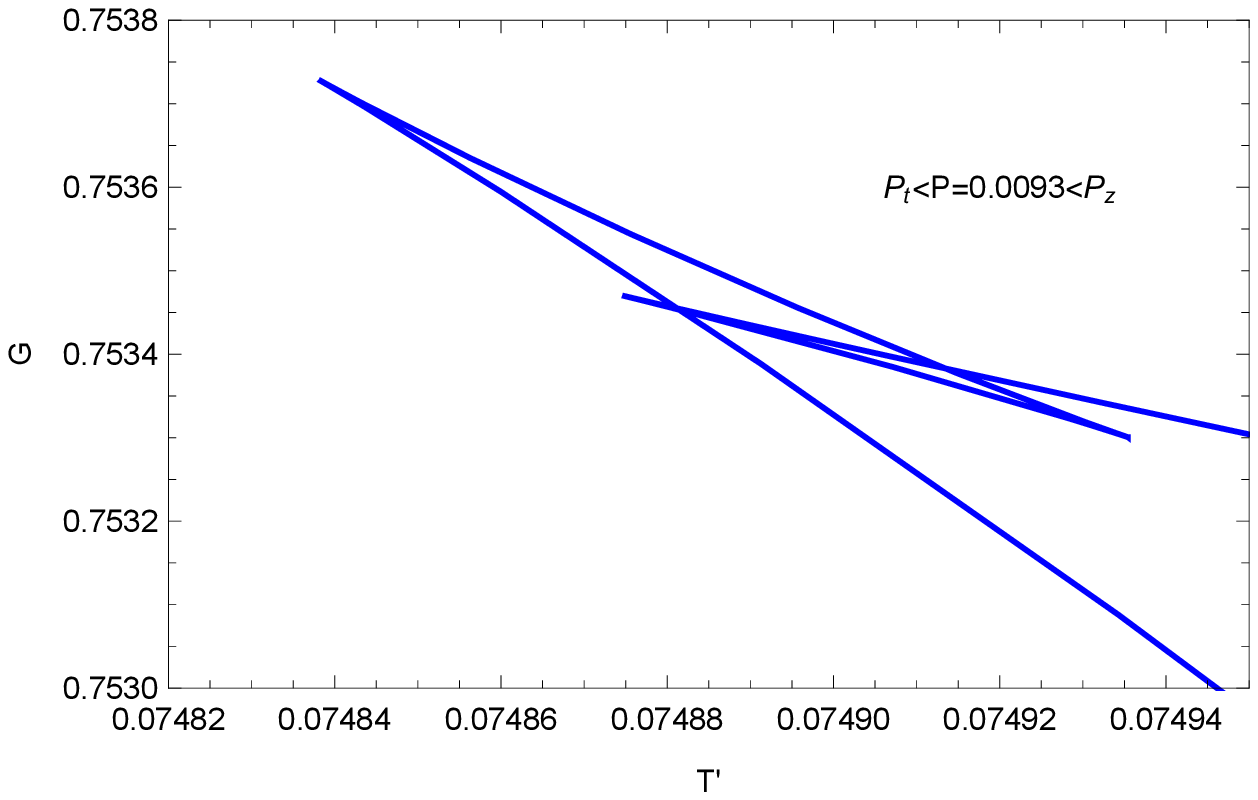}\hspace{0.5cm}
\includegraphics[width=5cm]{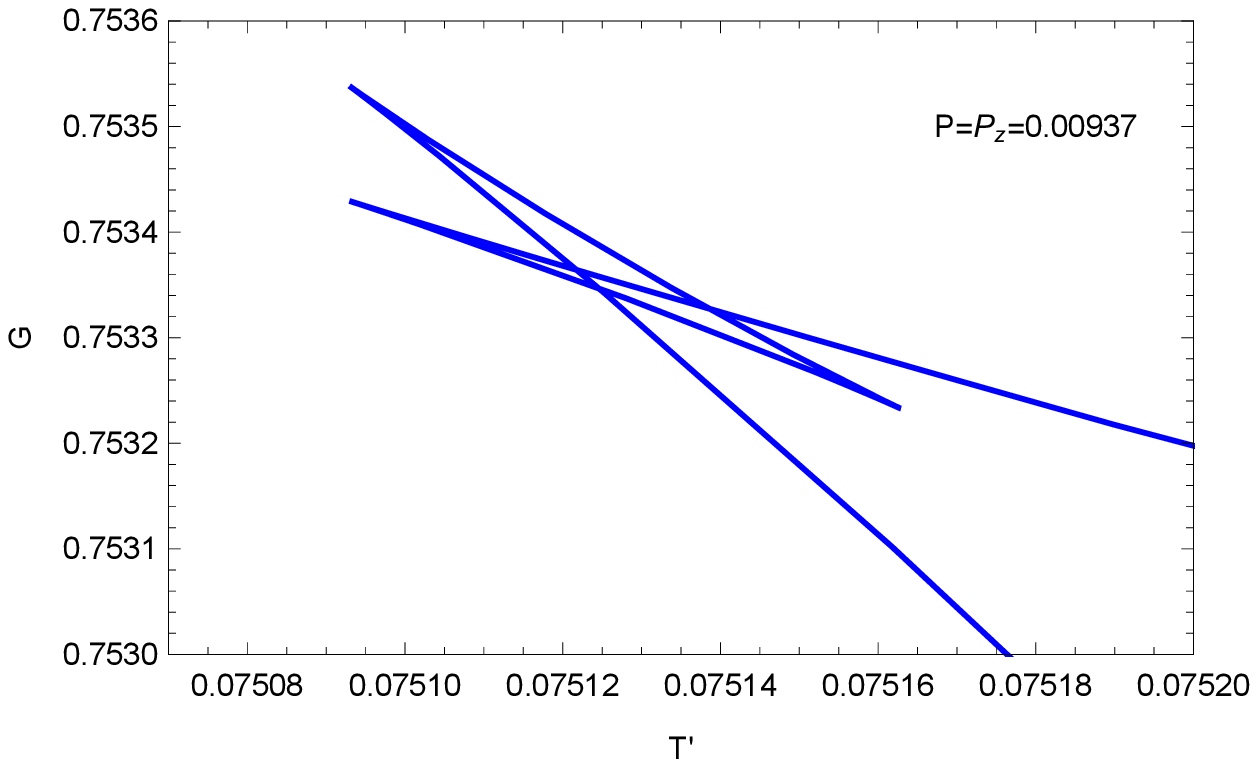}\hspace{0.5cm}
\includegraphics[width=5cm]{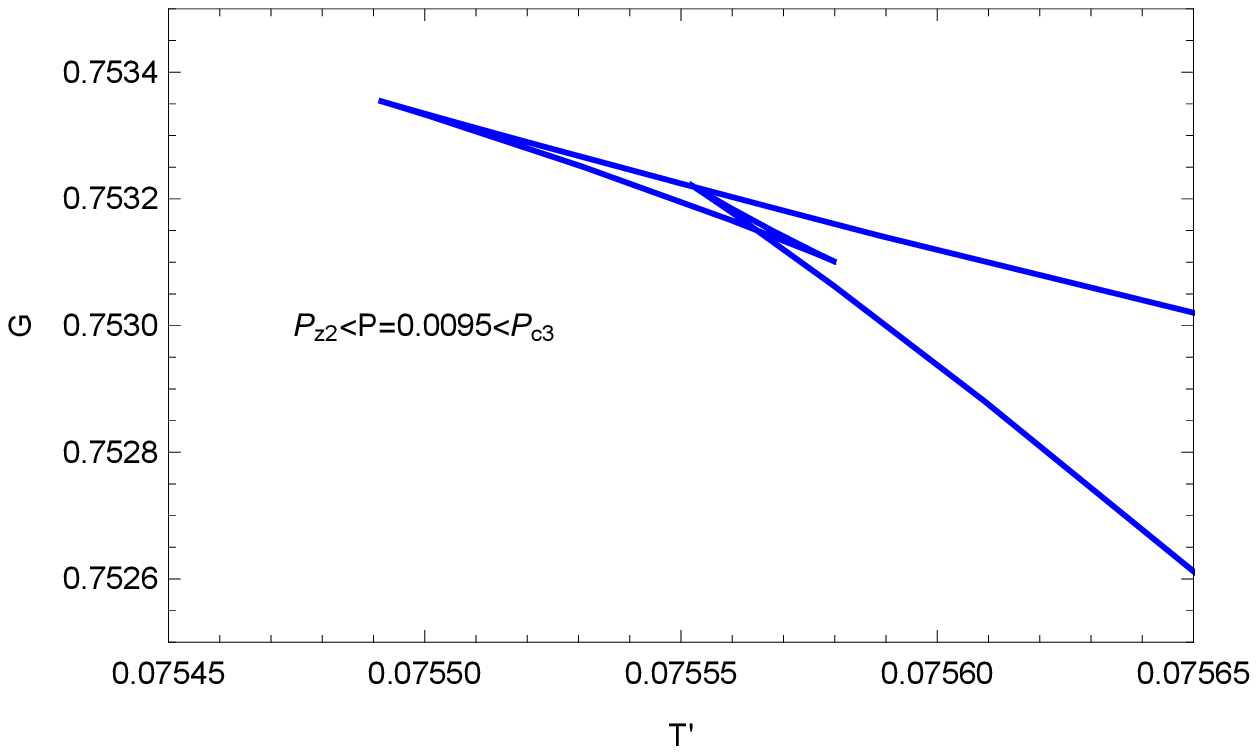}
\caption{The different behaviors of the Gibbs free energy at different pressures. $\alpha^2/Q^2=1.51$ with $\alpha=1$ and $Q=0.814$.}\label{fig32}}
\end{figure}

As is shown in Fig.\ref{fig31}, for $P<P_{c1}$ and $P\in (P_{c2},~P_{c3})$, there are four branches and there are two branches for $P>P_{c3}$ and $P\in (P_{c1},~P_{c2})$. According to the illustration in Fig.\ref{fig32}, when $P<P_{c1}$ and $P_{c2}<P<P_t$ there are swallowtail behaviors, however the swallowtail never intersects with the large black hole branch. When $P\in (P_{c1},~P_{c2})$ it is a cusp in the Gibbs free energy. This means that the large black hole is always globally thermodynamically stable when $P<P_{t}$ and $``c1"$ and $``c2"$ are not physical critical points. For $P>P_{c3}$, the Gibbs free energy also exhibits a cusp and for $P\in (P_t,~P_{c3})$ there is always a phase transition of first order. Thus, $``c3"$ is a physical critical point, which corresponds to a second-order phase transition.

\section{Conclusion and Discussion}
\label{Conclusions}

We begin by considering the charged AdS black hole with general topology ($k=0,\pm 1$). After considering the GUP we find that the temperature of the RN-AdS black hole, not like its semiclassical counterpart, can be always positive and has more fruitful structures. The GUP-corrected entropy is always positive and is independent of the electric charge $Q$ and the parameter $k$.
By analyzing the equation of state, we find that only in the $k=1$ case the black hole can have critical behaviors. In particular, one can judge the number of the critical points according to the values of $\alpha^2/Q^2$. Apparently, the number of the critical point can be one, two and three. In either case, there is the unique physical critical point. The main results have been summarized in Table.\ref{Tab1}.

For $P> 1.535$, the apparent critical point has a negative pressure, which is unphysical. At any positive pressure, the Gibbs free energy exhibits a cusp. For $P = 1.535$, the smaller one of the two apparent critical points also has negative pressure. Below or above the larger critical point, the $T'-r_h$ curve only has two branches, which corresponds to a cusp in the Gibbs free energy. For $1.5\leq P<1.535$, only the rightmost critical point is the physical one. In the first three cases in Table.\ref{Tab1}, there is always the VdW-like phase transitions, which occur at the range $P\in(P_t,~P_c$),~$P\in(P_t,~P_{c2})$, ~$P\in(P_t,~P_{c3})$ respectively and the RPT takes place for $P\in (P_t,~P_z)$.

\begin{table}[!hbp]
\centering
\begin{tabular}{|c|c|c|c| }
\hline\hline
 $\alpha^2/Q^2$ & $\#$ apparent critical points & $\#$ physical critical points & behavior \\
\hline
$(\frac{1}{1+8\pi P},~1.5)$ & 1 & 1 & VdW\& RPT \\
\hline
1.5  & 2 & 1 & VdW\& RPT \\
\hline
(1.5,~1.535)  & 3 &1& VdW\& RPT\\
\hline
1.535  & 2& 0 & cusp\\
\hline
$(1.535,~\infty)$& 1 & 0 & cusp\\
\hline
\end{tabular}
\caption{The critical behaviors of RN-AdS black hole with GUP corrections in the $k=1$ case.}\label{Tab1}
\end{table}

The Born-Infeld-AdS black hole and the Kerr-AdS black hole can have the reentrant behavior due to their complex horizon structure.
GUP only modify the thermodynamic quantities, but not the geometric structure like the metric. If taking into account the corrections of the GUP in these complicated black holes,
there should be more interesting phase structure and critical phenomena. This will be left for future studies.

\textbf{Conflicts of Interest}

The authors declare that they have no conflicts of interest.

\acknowledgments
This work is supported in part by the National Natural Science Foundation
of China (Grants No.11605107 and No. 11475108) and by the Natural Science Foundation of Shanxi (Grant No.201601D021022).

\bibliographystyle{JHEP}

\end{document}